# And again, about lasing threshold, light localization, and density of states

N.A. Vanyushkin, A.H. Gevorgyan

Far Eastern Federal University, 10 Ajax Bay, Russky Island, Vladivostok, 690922, Russia

**Abstract**

We investigated numerically the relationship between the lasing threshold, the density of states (DOS), and field localization in 1D photonic crystals (PC) for off-axis laser action. The angular dependence of the corresponding quantities for the two polarizations was obtained, and it was shown that the relationship between the threshold and DOS is complex, especially near the Brewster angle in the case of p-polarization. A strong correlation between the ratio of thresholds and localizations of the two edge modes was found for s-polarization. The results of this work can be used for optimizing lasers based on one-dimensional PCs, both for normal and off-axis lasing.

1. Introduction

Lasers have found a wide range of applications in almost all fields of science and technology. The first lasers were created back in the 60s of the last century, but they continue to attract considerable attention of researchers to this day. The main classes of lasers include gas lasers, solid-state lasers, dye lasers, and semiconductor lasers [1], each of which can be subdivided into a number of subclasses with their own specific features.

A special position is occupied by semiconductor lasers with distributed feedback (DFB). The research and development of semiconductor DFB lasers began quite a long time ago [2]. Such a laser has a periodic change in the refractive index along the length of the active region. The small size, extremely narrow and adjustable output spectrum, as well as the mature technology of semiconductor structures production have allowed DFB lasers to become an indispensable element of modern optical telecommunications. The laser operation in a medium with distributed feedback, which also includes photonic crystals (PCs), has a number of unique features that distinguish it from the lasing in conventional Fabry-Perot laser resonators. PC-based lasers, as a rule, are characterized by small size, low generation threshold, small number of longitudinal modes generated, and, as a result, a narrow spectrum of the output radiation. In addition, the phenomenon of off-axis laser generation allows simultaneous generation at many wavelengths.

Among the studies related to laser generation in PC, of particular interest are the reduction of the generation threshold and increasing its efficiency. In this case, finding numerically the lasing threshold itself for an arbitrary structure is a relatively time-consuming task, since generally it requires solving a boundary problem for the field amplitude. For this reason, a common practice in the analysis of generation in PCs is the study of the optical density of states (DOS), which is much more convenient for work than the direct calculation of the threshold [3-6]. In the literature, DOS is usually understood as the "number of states" with different wave vectors $k$ in a unit spectral interval $d\omega$, which can be represented as $\rho = dk/d\omega$. It follows from this definition that the DOS is the inverse of the group velocity in the medium $v_g = 1/\rho$. It is worth noting that the

DOS of PCs has a rather complex dependence on the structure parameters, and therefore is usually determined numerically like the threshold. Still, approximate dependences on some parameters are known, for example, the DOS at edge modes depends on the number of unit cells $N$ as $\rho \sim N^2$, while the DOS inside the photonic band gap (PBG) decreases as $\rho \sim 1/N$ [5]. Numerous works have shown a good correlation between threshold and DOS in various types of resonators, including PCs (see, for example, [7-10]). Nevertheless, only in a few cases, for example, for a homogeneous Fabry-Perot resonator, it is possible to obtain a clear relation between these two quantities [6]. In other cases, the absence of such a relation between the DOS and the lasing threshold significantly complicates the problem of laser optimization.

One promising way to further lower the lasing threshold is the use of off-axis generation on conical modes [11-14], when the laser action occurs along directions that are not parallel to the PC axis. The set of such directions forms a system of concentric cones with the axis coinciding with the axis of the PC. For this reason, off-axis laser generation is also called conic, and spatial modes of output radiation are called conical modes. One of the features of off-axis generation is that each conical mode has its own distinct wavelength, and the larger the cone angle, the shorter the wavelength. Under certain conditions it is possible to observe lasing of many conical modes at once, which opens the possibility of obtaining a compact multi-frequency laser. Off-axis lasing has been obtained both in conventional solid-state active PCs [13,14] and in lasers on cholesteric liquid crystals [11,12].

In this paper we investigated the relationship of the generation threshold in binary PC with DOS and light localization depending on angle of incidence.

## 2. Theory

Let us consider the reflection and transmission of a plane electromagnetic wave through a 1D binary PC (Fig. 1) which consists of $N$ unit cells and is located between the planes $z = 0$ and $z = L$, and its refractive index $n$ depends on $z$ only. We assume that all layers are isotropic, nonmagnetic ($\mu = 1$), and do not absorb light. Let us also assume that the incidence plane coincides with the plane $(x, z)$, and the wave is incident at an angle $\alpha_0$ to the normal of the PC boundary, which coincides with the plane $(x, y)$. The regions $z < 0$ and $z > L$ are filled with homogeneous non-absorbing dielectric with refractive index $n_0$. Complex refractive indexes of both layers in the unit cell have the following form:

$$\hat{n}_{1,2} = n_{1,2} - i\frac{c}{\omega}g_{1,2} \qquad (1)$$

Here $g$ is the gain coefficient, which is supposed to be constant for all frequencies.

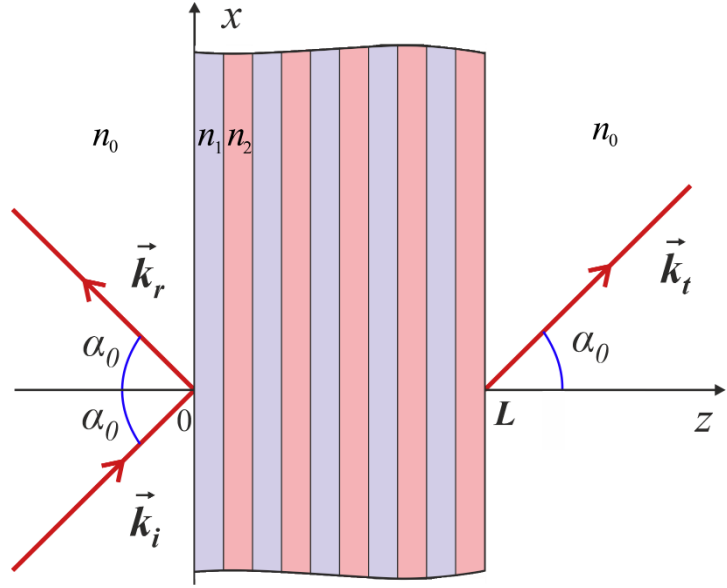

Figure 1. Geometry of the problem.

When solving this problem, we employed the transfer matrix approach [15,16] to calculate the transmission spectra of the PC. The transfer matrix relates the field amplitudes at different interfaces and can be presented for $j$-th layer in the following form:

$$M_j = \begin{pmatrix} \cos k_j d_j & -\dfrac{i}{p_j}\sin k_j d_j \\ -i p_j \sin k_j d_j & \cos k_j d_j \end{pmatrix}, \qquad (2)$$

where $k_j = (\omega/c) n_j \cos\alpha_j$, $d_j$ is the thickness of the layer, $\alpha_j$ is the angle of refraction which is determined by $n_j \sin\alpha_j = n_0 \sin\alpha_0$. Also $p_j = n_j \cos\alpha_j$ for s-wave and $p_j = (1/n_j)\cos\alpha_j$ for p-wave. Then the transfer matrix $m$ of entire PC can be obtained by multiplying the transfer matrices of all layers:

$$m = (M_1 M_2)^N = \begin{pmatrix} m_{11} & m_{12} \\ m_{21} & m_{22} \end{pmatrix} \qquad (3)$$

Carrying out some mathematics then the transmission coefficient for s- and p-waves are given as

$$t = \frac{2 p_0}{(m_{11} + m_{12} p_0) p_0 + (m_{21} + m_{22} p_0)} \qquad (4)$$

Here $p_0 = n_0 \cos\alpha_0$ for s-wave and $p_0 = (1/n_0)\cos\alpha_0$ for p-wave. The corresponding energy coefficient is calculated as $T = |t|^2$.

There exist several ways to calculate DOS which are described and compared in detail in [3]. Weighted averaging of local density of states (LDOS), which in turn obtained via the Green's function, is considered the most general way to calculate DOS of a structure. However, in many cases it is more convenient to calculate it via so called phase time (or Wigner time) $\tau_\varphi$ which requires only spectral dependence of transmission coefficient $t$:

$$\rho = \frac{\tau_\varphi}{L} = \frac{1}{L}\frac{d}{d\omega}(Arg(t)) = \frac{1}{L}\operatorname{Im}\left(\frac{1}{t}\frac{dt}{d\omega}\right) \qquad (5)$$

Below we will use DOS $\rho$ normalized to its value $\rho_0 = \langle n \rangle / c$ for homogeneous medium with mean refractive index $\langle n \rangle$ of the PC under consideration so that $\rho$ for such media will be equal to unity. Also, in this paper when calculating DOS for the PC we will assume absence of amplification ($g_1 = g_2 = 0$).

Further, we will consider our PC to reach the lasing threshold when the denominator of Eq. (4) turns into zero, i.e. the transmission reaches infinity [16]. This condition can be interpreted as when lasing is achieved one can see a finite outcoming wave in the absence of any incoming wave. So, if the condition $1/t(\lambda, g) = 0$ is satisfied for some pair of parameters $(\lambda_0, g_{th})$, then the lasing is achieved at the wavelength $\lambda_0$ with the lasing threshold $g_{th}$. It is also worth noting that the lasing can be achieved only at specific wavelengths while at all other wavelengths the lasing is impossible no matter how much the amplification is.

### 3. Numerical simulations and discussion

One of the features of binary PCs is that we can control the DOS spectrum by changing the ratio of the optical thicknesses of the layers $n_1 d_1 / n_2 d_2$ in the unit cell. The PC with $n_1 d_1 / n_2 d_2 = 1$ is of particular interest and is called a quarter-wave stack. Such a PC has a number of peculiarities: 1) absence of PBGs of even orders; 2) maximum PBG width is achieved for a given value of refractive index contrast ($\Delta n = |n_1 - n_2|$); 3) symmetry of transmission $t$ and DOS spectra relative to the PBG center [4]. Figure 2 shows the DOS spectra near the first PBG of our PC with different values of the ratio $n_1 d_1 / n_2 d_2$. It is also worth noting that each DOS maximum corresponds to its mode with transmission $T = 1$, although the transmission and DOS maxima do not coincide exactly with each other. Further, we will assign positive numbers (+1, +2, +3...) to the short-wave modes and negative numbers (-1, -2, -3...) to the long-wave modes, where the modes -1 and +1 are nearest to the PBG.

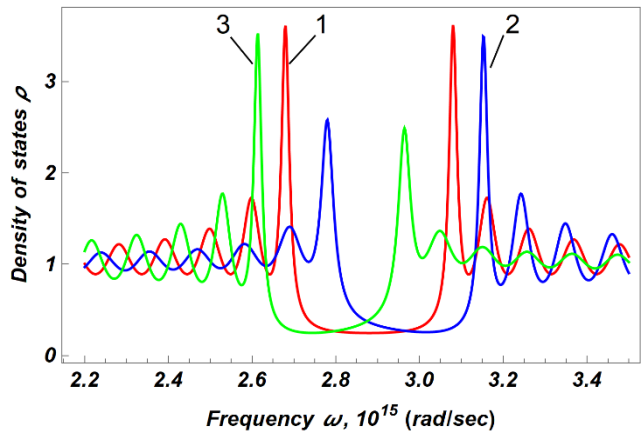

Figure 2. DOS spectra of binary PCs with different $n_1 d_1 / n_2 d_2$ ratio: curve 1 $n_1 d_1 / n_2 d_2 = 1$, curve 2 $n_1 d_1 / n_2 d_2 = 2$, curve 3 $n_1 d_1 / n_2 d_2 = 0.5$. Parameters of the structure: $n_1 = 1.5$, $n_2 = 1.8$, $d_1 + d_2 = 200$ nm, number of unit cells $N = 25$, $n_0 = 1$, normal incidence.

### 3.1. The influence of spatial amplification distribution on lasing threshold

When investigating laser generation in layered PCs, first of all it is necessary to choose the distribution of the gain along the length of the PC, namely, which layers will be active (amplifying). Let us consider three main cases for our PC: 1) $g_1 \neq 0$ and $g_2 = 0$, 2) $g_1 = 0$ and $g_2 \neq 0$, 3)

$g_1 = g_2 = g \neq 0$. Figure 3a shows the lasing thresholds for several modes as well as the spectrum of DOS in the vicinity of the first PBG. We can see that, in the first case, the modes to the left of the PBG (long-wave) reach the threshold earlier than the modes to the right of the PBG (short-wave). In the second and third cases, the situation is exactly the opposite.

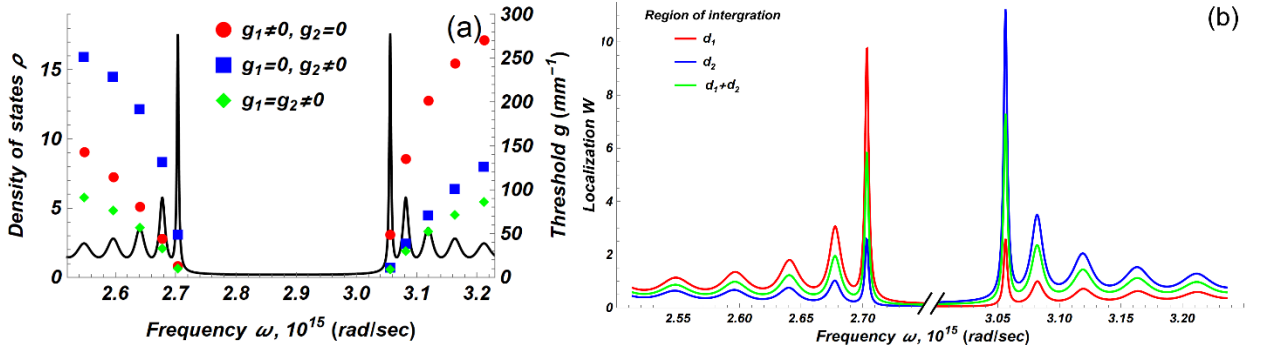

Figure 3. (a) DOS spectrum of quarter-stack PC and lasing thresholds for different modes. (b) Field localization spectra comparison. Parameters of the structure: $n_1 = 1.5$, $n_2 = 1.8$, $d_1 + d_2 = 200$ nm, $n_1 d_1 / n_2 d_2 = 1$, number of unit cells $N = 50$, $n_0 = 1$, normal incidence.

This observation agrees well with the corresponding differences in the localization of the mode field to the right and left of the PBG, which is shown in Figure 3b. The field localization was calculated using the following formula:

$$W = \int_S |E|^2 \, dz / \int_S dz, \tag{6}$$

where the region of integration $S$ corresponds to the layers with $n_1$, the layers with $n_2$, or the entire length of the PC (from $z = 0$ to $z = L$).

For modes to the left of the PBG, the field is mainly localized inside layers with the lower refractive index ($n_1$), while modes to the right of the PBG, on the contrary, are localized in layers with the higher refractive index ($n_2$). This pattern is not limited to binary PCs and is also observed under other refractive index dependences [17]. Thus, the lasing threshold inversely correlates with the integral of the overlap between the wave field and the gain coefficient when the DOS is constant. In addition, we empirically found that the lasing thresholds of edge modes of the same order (±1, ±2, etc.) practically coincide at the gain ratios $g_1 = n_1 g$ and $g_2 = n_2 g$, that is $g_1 / g_2 = n_1 / n_2$. It is this ratio of gain coefficients that we have chosen for further study, in order to minimize the effect of field localization within a certain layer on the lasing threshold at normal incidence.

### 3.2. Angular dependence of DOS and lasing threshold

Now let us go directly to the study of dependence of DOS and lasing threshold on the polarization of radiation and the incidence angle $\alpha_0$. Figures 3a,b show obtained dependences for DOS and lasing threshold of our PC on the incidence angle for long-wave and short-wave edge modes (orders -1 and +1, respectively) for both polarizations in the case when refractive index of external medium $n_0 = 1$. As the incidence angle increases, the DOS of both modes for the s-wave increases and the thresholds decrease, while the opposite is true for the p-wave at angles smaller than the Brewster angle. As mentioned above, at normal incidence both DOS and thresholds coincide for the two edge modes. As the incidence angle increases, both quantities start to diverge. The difference in DOS between the two modes in Figures 4a,b can be explained by the fact that the quarter-wave condition for the unit cell is violated, since the effective optical thickness $d_j n_j \cos\alpha_j$ can be equal for the two layers only at one angle of incidence (in our case it is $\alpha_0 = 0$). The inevitable change in the ratio $d_1 n_1 \cos\alpha_1 / (d_2 n_2 \cos\alpha_2)$ at an oblique incidence leads to an asymmetry in the DOS distribution with respect to the PBG, similar to that in Figure 2, which can partially explain the differences in the lasing thresholds of the two modes. Nevertheless, the DOS of the two modes start to differ noticeably only at incidence angles greater than 65 degrees, while the thresholds already diverge at 20 degrees, indicating the influence of some other mechanism. Finally, Figures 4a,b compare the DOS and threshold values with each other. For the s-polarization, the dependence is close to $g \sim 1/\rho$, which agrees well with the result for the Fabry-Perot resonator [6]. For p-polarization, a more complex dependence between these quantities is observed, especially near the Brewster angle. To summarize, we can confidently say that laser generation with s-polarization dominates at all angles, which results in a good extinction ratio of the output radiation of such a laser.

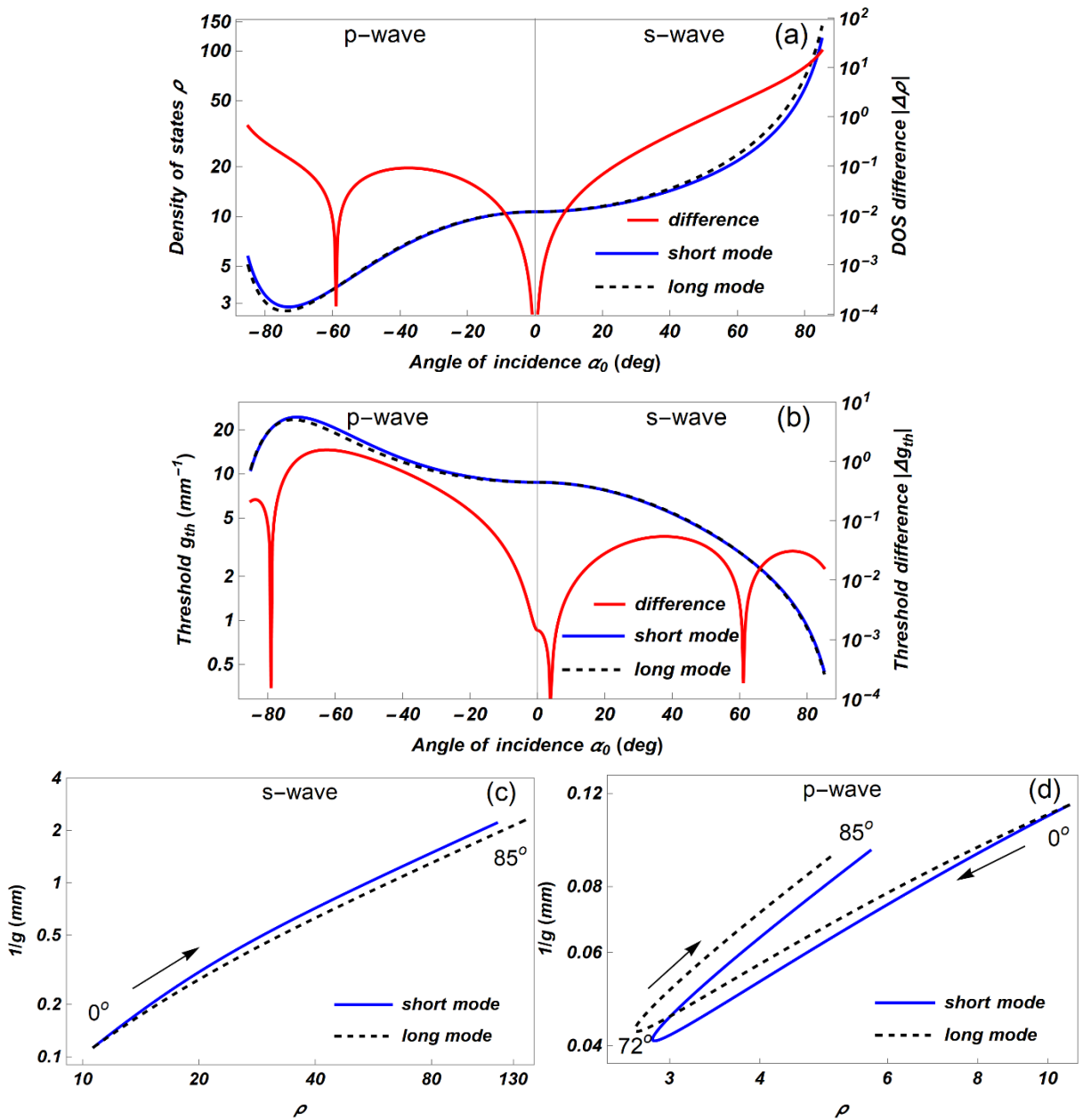

Figure 4. Dependences of DOS and lasing threshold on the angle of incidence for s-polarization and p-polarization at $n_0 = 1$. $g_1 / g_2 = n_1 / n_2$, the other parameters are the same as in Fig. 3.

Next, we will consider the same dependences when $n_0$ equals to the average refractive index of the PC $\langle n \rangle$. This value of $n_0$ allows to minimize the reflection from the PC boundaries, and also allows to increase the maximum achievable refractive angle inside the PC layers. The corresponding graphs are shown in Figure 5. In general, the obtained dependences are similar to those in the previous case, but there are a number of notable differences.

First, the DOS of the two edge modes practically coincide in a range of angles up to about 60 degrees, after which a sharp divergence begins. At large incidence angles the angle of total internal reflection $\cos\alpha_1 = 0$ for the layer with $n_1$, which means that the relation $d_1 n_1 \cos\alpha_1 / (d_2 n_2 \cos\alpha_2)$ also turns zero, and an extremely large asymmetry of the DOS distribution with respect to the PBG appears. In our case, the angle of total internal reflection is reached at $\alpha_0 = 66.4$ degrees, which is accompanied by a significant difference in DOS, and at $\alpha_0 = 60.8$ degrees $d_1 n_1 \cos\alpha_1 / (d_2 n_2 \cos\alpha_2) = 1/2$, which is a characteristic angle when the optical thicknesses of the layers begin to differ noticeably, and as can be seen, at this angle the DOS of the two modes begins to diverge considerably.

Second, at large angles of incidence, the DOS and the lasing threshold of the p-polarization practically match the corresponding values for the s-polarization, which leads to competition between polarizations and deterioration of the extinction ratio of the laser output radiation.

Third, Figures 5c,d show a noticeable deviation of the dependence $g \sim 1/\rho$ for the short-wavelength mode at very large incidence angles (near the DOS maximum), which again appears to be related again to reaching the angle of total internal reflection.

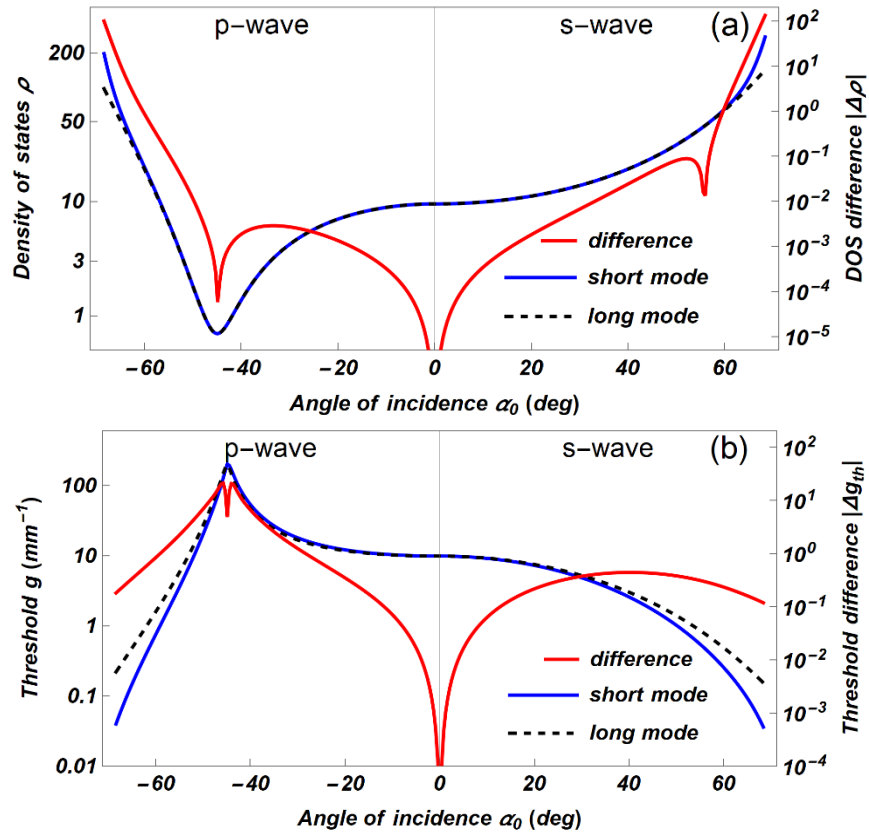

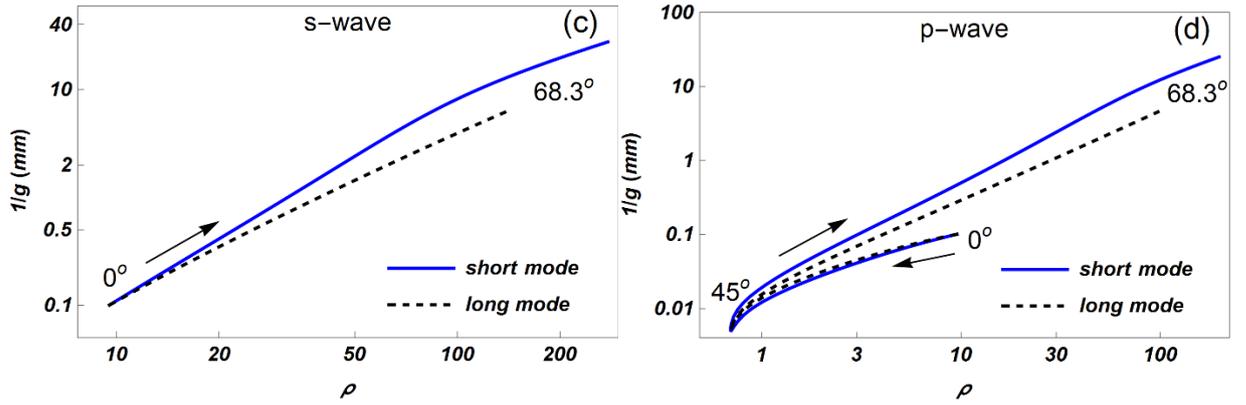

Figure 5. Dependences of DOS and lasing threshold on the angle of incidence for s-polarization and p-polarization at $n_0 = \langle n \rangle$. $g_1 / g_2 = n_1 / n_2$, the other parameters are the same as in Fig. 3.

In addition, it is worth noting that often in practice the lasing threshold of conical modes turns out to be larger than for the normal mode. In our model, we consider an infinite in transverse dimensions PC with a homogeneous gain in each individual layer, which has a positive effect on off-axis laser action. In reality, all one-dimensional PCs have a finite transverse area, and the well-pumped transverse area of a PC with inverse population (high gain) is even smaller. In such situation the normal mode may have an advantage over the conical modes, and therefore have a lower generation threshold. For this reason, it is important to choose the right transverse size and pumping scheme of the laser in order to take full advantage of off-axis lasing.

### 3.3. Light localization and lasing threshold

Above we have again seen that the DOS alone is not enough for an accurate prediction of the threshold. One of the other parameters characterizing the lasing threshold is the localization of the field (6) inside the structure under study. In addition to the usual localization calculated by formula (6), it may be more advantageous to use the weighted field localization with the gain as a weight function:

$$W_{adj} = \int_0^L g(z)|E|^2 \, dz \bigg/ \int_0^L g(z) \, dz \qquad (7)$$

In our opinion, the weighted field localization better accounts for the influence of the spatial distribution of the gain on the lasing threshold.

Figure 6 shows the change in the ratio of various quantities (field localization, threshold, and DOS) from the angle for two first-order edge modes (+1 and 1), which are normalized to the corresponding values at normal incidence ($\alpha_0 = 0$). First of all, it is worth noting the significant difference in the "threshold - DOS" and "threshold - localization" correlations in the case of s-polarization. It seems that the DOS ratio of the two modes has no significant effect on the threshold ratio, while the relative change in threshold and localization is practically the same for all incidence angles. Thus, the angular dependence of lasing thresholds of two modes for s-polarization can be expressed through field localization:

$$\frac{g_{long}(\alpha_0) W_{long}(\alpha_0)}{g_{short}(\alpha_0) W_{short}(\alpha_0)} \approx \frac{g_{long}(0) W_{long}(0)}{g_{short}(0) W_{short}(0)} \qquad (8)$$

We found that relation (8) is fulfilled with good accuracy not only at $g_{long}(0) = g_{short}(0)$ and $\rho_{long}(0) = \rho_{short}(0)$, as in our case, but also at other relations of initial lasing thresholds and DOS. This fact makes it easy to calculate the angular dependence of one mode, if such dependence is known for another mode of the same order and the field localization dependence for both modes, which can be useful both in numerical simulation of lasing in PC, and in theoretical analysis. It is not necessary to know the DOS of any mode to do so. It is also worth noting that when using weighted localization (7), relation (8) is satisfied somewhat better than when using conventional field localization (6), as can be seen from Figures 6a,c. Thus, we can say that in this case it is the field localization that determines the difference in the lasing thresholds of the two modes. In the case of p-polarization such a good correlation between the threshold and field localization is no longer observed. Moreover, we can see that the field localization and DOS correlate with each other even better than with the lasing threshold.

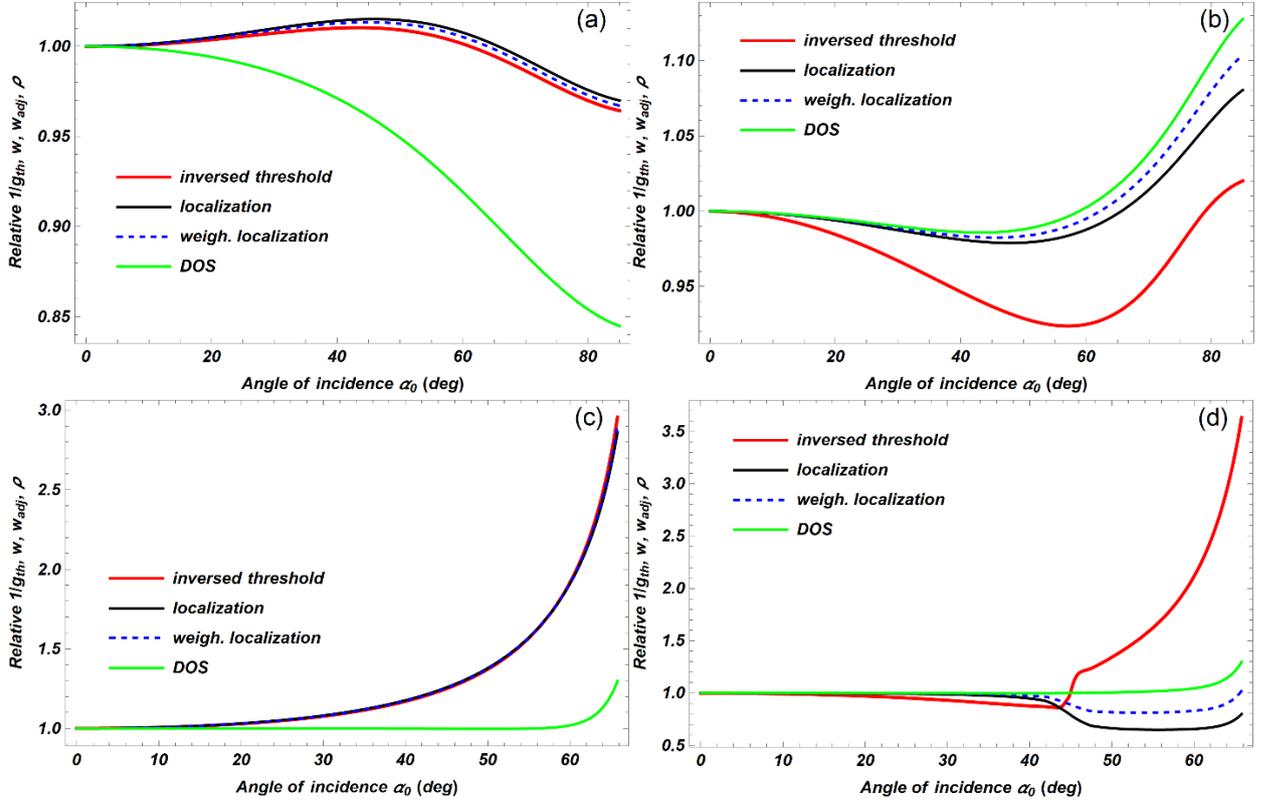

Figure 6. Angular dependences of ratio of field localization $W_{short}/W_{long}$ and $W_{short}^{adj}/W_{long}^{adj}$, DOS $\rho_{short}/\rho_{long}$ and lasing threshold $g_{long}/g_{short}$ between shortwave and longwave edge modes for s- and p-polarization (a,b) at $n_0 = 1$ and (c,d) at $n_0 = \langle n \rangle$. $g_1/g_2 = n_1/n_2$, the other parameters are the same as in Fig. 3.

## 4. Conclusions

In conclusion, in this work we investigated the relationship between DOS and light localization with the lasing threshold for binary PC at different angles of incidence. It was shown that differences in field localization within the layers with higher or lower refractive indices have a significant effect on the generation threshold between the longwave and shortwaves modes (relative to the PBG), even when the DOS is the same. At the same time the ratio between gain of two layers was found $g_1/g_2 = n_1/n_2$, at which the lasing thresholds of two modes of the same order are equal. Further the relation between DOS and threshold at off-axis lasing was investigated.

It was shown that for s-polarization the dependence is close to g ~ $1/\rho$, similar to the case of normal incidence in the Fabry-Perot resonator at $\rho \gg 1$, while for p-polarization the dependence has a more complex form, especially near the Brewster angle. Moreover, for most angles of incidence (the cone angle), the lasing threshold for s-modes is much smaller than for p-modes. The difference between polarizations is especially strong again near the Brewster angle, which can be used, for example, to obtain conical modes with stable polarization and high extinction ratio. Finally, the angular dependence of the DOS ratio, localization, and generation threshold between two first-order edge modes was investigated. A strong correlation between the lasing thresholds and field localization of the two modes for s-polarization was found. This fact allows to easily calculate the angular dependence of one mode, if such dependence is known for another mode of the same order and the field localization dependence for both modes, which can be useful both in numerical simulation of laser generation in FC, and in theoretical analysis. In addition, the results of this work can be used in the development of methods for optimization of lasers based on one-dimensional PCs, both for normal and for off-axis lasing.

## Acknowledgments

The work was supported by the Foundation for the Advancement of Theoretical Physics and Mathematics "BASIS" (Grant № 21-1-1-6-1).

## Author agreement

We declare that this manuscript is original, has not been published before and is not currently being considered for publication elsewhere. We confirm that the manuscript has been read and approved by all named authors and that there are no other persons who satisfied the criteria for authorship but are not listed. We further confirm that the order of authors listed in the manuscript has been approved by all of us. We understand that the Corresponding Author is the sole contact for the Editorial process. He is responsible for communicating with the other authors about progress, submissions of revisions and final approval of proofs.

## Declaration of competing interest

The authors declare that they have no known competing financial interests or personal relationships that could have appeared to influence the work reported in this paper.

## CRediT authorship contribution statement

A.H. Gevorgyan proposed the idea and contributed to the data analysis. N.A. Vanyushkin developed procedures for investigating and wrote the manuscript, contributed to the Software, to the Data curation and to the data analysis. All authors contributed to the discussion of this work, to the Validation, Writing - review & editing.

## Conflict of interest

No potential conflict of interest was reported by the authors.